\documentclass[a4paper,11pt]{article}
\usepackage{jcappub} 
\usepackage{lineno}
\usepackage{multirow}

\title{\boldmath Searching for QPOs in BATSE Short Gamma-ray Bursts Based on Narrowband and Broadband Features}

\author{Dong-Jie Liu and Yuan-Chuan Zou}
\affiliation{Department of Astronomy, School of Physics, Huazhong University of Science and Technology,
Wuhan, 430074, China}

\emailAdd{zouyc@hust.edu.cn}

\abstract{Gamma-ray bursts (GRBs), especially short GRBs, are often considered potential candidates for exhibiting kilohertz quasi-periodic oscillations (QPOs) due to their origin from binary mergers. It has already been discovered that two bursts exhibit QPOs. While systematic searches for QPOs in GRBs typically concentrate on the kilohertz range, there has been no comprehensive exploration in the hundred-hertz range. In this study, we systematically conducted QPO searches on all BATSE short burst data within the 0-1000 Hz range. Using nested significance tests, we observed that the reference distributions for different GRBs are quite similar. This observation prompted us to analyze the data by selectively focusing on those with larger statistical values, obviating the need to iterate through all the data and significantly reducing computational workload. Ultimately, our findings did not reveal any compelling evidence for QPOs, which may suggest that the GRB jet has lost the early merging memory.}

\begin{document}
\maketitle
\flushbottom

\section{Introduction}
\label{sec:intro}

Gamma-ray bursts (GRBs) are among the most powerful explosions in the universe. On the basis of their duration, GRBs can be classified into long GRBs (LGRBs) and short GRBs (SGRBs). It is generally believed that LGRBs originate from the collapse of massive stars, while SGRBs result from the merger of compact binary systems (binary neutron stars or neutron star-black hole)~\cite{Piran:2004ba,Kumar:2014upa}. The association between GW 170817 and GRB 170817A indicates that at least a subset of short bursts originates from the merger of binary neutron stars~\cite{LIGOScientific:2016aoc,LIGOScientific:2017zic}. 

In astronomical observations, the search for periodic or quasi-periodic signals has been a continual focus of research. Several studies have previously attempted to detect such signals in LGRBs or SGRBs, and even in the precursor stages~\cite{Kruger:2001ti,Guidorzi:2016ddt,Tarnopolski:2021ula,Xiao:2022tqy}. A rapid Fourier transform was conducted by~\cite{1997ApJ...491..720D} on 20 bright GRBs, searching for periods within the range of 0.016 to 33.3 milliseconds, but no significant evidence was found. Another study by~\cite{Kruger:2001ti} utilized all Time-Tagged Event (TTE) data from BATSE, covering 2203 GRBs and 152 Soft Gamma-ray Repeaters (SGRs), employing Rayleigh power as an indicator to search for periodic signals and also found no evidence. Subsequent investigations used wavelet analysis for periodic signal searches~\cite{2009AN....330..404Z}. Furthermore, a study using a Bayesian framework for period searches found no evidence for quasi-periodic oscillations (QPOs) in 44 bright SGRBs~\cite{Dichiara:2013dpa}.
In a recent study, QPOs in the kilohertz range were identified in two bursts within the BATSE SGRB data~\cite{Chirenti:2023dzl}. Additionally, there has been work suggesting the presence of quasi-periodic signals in GRB 211211A~\cite{Xiao:2022quv,Chirenti:2023cgk}, which originates from a kilonova.

The most common method for detecting periodicity is through power spectra or periodograms, involving the transforming of time-domain information into the frequency domain via Fourier transformation. Based on the assumption that the numerical values within each interval follow a $\chi_2^2$ distribution~\cite{2005A&A...431..391V}, i.e., $\chi^2$ in distribution with 2 degrees of freedom, if the value at a specific frequency significantly deviates from this distribution, it is considered a potential indication of a periodic signal.
Empirical formulas are typically used to characterize the profile of observational data, and noise is then added to this profile. Monte Carlo simulations are employed to generate simulated signals with added noise. By comparing the observed signal with the simulated signal, one can assess whether the observational data contains potential periodic signals. Two distinct scenarios exist: simulating light curves and simulating periodograms.
For GRBs, when simulating light curves, Gaussian or fast rise exponential decay (FRED) functions are commonly used to characterize the profile~\cite{Kocevski:2003ae,Norris:2005ew}. Poisson noise is then added to this profile. When simulating periodograms, the profile is often represented using power law (PL) or broken power law (BPL) functions~\cite{2010MNRAS.402..307V}, and the values at each frequency are perturbed using a $\chi_2^2$ distribution. Additionally, the traditional approach involves fitting simulated signals onto the optimal profile, whereas Bayesian methods utilize Markov Chain Monte Carlo (MCMC) to randomly sample from the posterior distribution, followed by the addition of noise.~\cite{Huppenkothen:2012am} pointed out that searching for periods and searching for QPOs should involve different methods. Periods often refer to a narrow peak on the periodogram, while QPOs allow for fluctuations in the period within a small range, presenting a broad peak on the periodogram. These are narrowband and broadband features. They employed two different methods to capture these two features in the periodogram~\cite{Huppenkothen:2012am,Huppenkothen:2014pba}. In this paper, we adopt their methods for our analysis.

While there have been some systematic searches for QPOs in short bursts, the statistical analysis has generally focused on frequencies above 400Hz or 500Hz. In the low-frequency range, searches have been conducted for only a subset of bright GRBs, and a comprehensive systematic search has not been performed yet. On the one hand, this is because the QPOs predicted by the model mostly occur in the kilohertz range. On the other hand, the influence of red noise in the low-frequency range is significant.

Our study is dedicated to systematically searching for QPOs in BATSE SGRB TTE data across the frequency range of 0-1000 Hz.   Our approach involves both narrowband and wideband analyses. In the narrowband analysis, we scrutinize all accessible SGRBs from BATSE, while in the wideband analysis, we initially narrow down the selection to a subset of GRBs to address computational constraints. The paper is structured as follows. Section 2 describes our data selection criteria and analysis methods. Section 3 presents the results of our investigation. Finally, Section 4 delves into the discussion and provides a summary.

\section{Data Sample and Method}

\subsection{Data Selection}

Our initial data set is sourced from the Time-Tagged Event (TTE) data of 532 SGRBs obtained from the Compton Gamma Ray Observatory (CGRO)/BATSE\footnote{\url{https://heasarc.gsfc.nasa.gov/FTP/compton/data/batse/ascii_data/batse_tte/}}~\cite{Preece:1999fv}. The time resolution is 2$\mu s$. 
The detector consists of four channels, covering energy ranges of 20 to 50 keV, 50 to 100 keV, 100 to 300 keV, and above 300 keV, respectively. In this regard, we excluded data from seven bursts with the following identifiers: 298, 603, 878, 2103, 2288, 2988, and 6293. The main reasons for exclusion were poor data quality or the presence of potential issues. Additionally, in pulses 2103 and 2988, we detected a significantly anomalous QPO around 900 Hz in the later stages, raising suspicions of instrumental problems.

\subsection{Method}
We used a Bayesian method based on the periodogram to search for QPO signals~\cite{Huppenkothen:2012am}. By employing fast Fourier transform and Leahy normalization~\cite{1983ApJ...266..160L}, we obtained the periodogram. The periodogram can be modeled as a combination of red noise at low frequencies and white noise at high frequencies. The white noise component can be considered as a constant term, whereas the red noise component can be modeled using a PL or BPL model. Therefore, the problem can be framed as a model selection issue between the two models. One is the power law plus a constant (PLC),
\begin{equation}
\label{eq1}
P(\nu) = \beta\nu^{-\alpha} + \gamma,
\end{equation}
and the other is a broken power law plus a constant (BPLC),
\begin{equation}
\label{eq2}
P(\nu) = \beta\nu^{-\alpha_1}\left[1 + \left(\frac{\nu}{\delta}\right)^{(\alpha_2 - \alpha_1)/\rho}\right]^{-\rho} + \gamma.
\end{equation}
In both models, $\beta$ is a normalization term and $\gamma$ is a constant term that represents white noise. $\alpha$ is the PL model index, $\alpha_1$ and $\alpha_2$ represent the indices of the BPL model at low and high frequencies, respectively. $\delta$ is the breaking frequency between the two power-law components. $\rho$ is a smoothness parameter, and here we set $\rho$ = 1 to better constrain the other parameters. The maximum a posteriori (MAP) method is used as a parameter estimation method. Then, we can obtain the likelihood ratio between the two models. The likelihood ratio test (LRT) statistic is commonly used for model selection. Using an MCMC technique~\cite{Metropolis:1953am,Hastings:1970aa}, we sample the posterior predictive distribution and generate simulated periodograms with no periodicity. For each simulated periodogram, we can obtain its likelihood ratio through the above steps. Thus, we get a distribution of likelihood ratios for simulated periodograms. Placing the obtained likelihood ratio from the real periodogram into this distribution. If it significantly deviates from the distribution, the alternative hypothesis is favored over the null hypothesis. For the noise model, we choose the alternative model when the posterior predictive $p$-value is less than 0.05. Once the noise model is selected, we can search for QPOs from both narrowband and broadband features.

To investigate the features of narrowband, the test statistic $T_R = {\rm max}_j(2I_j/S_j)$ is used, where $I_j$ and $S_j$ are observed and model powers at frequency $j$. The residual $R_j = 2I_j/S_j$ is commonly used in the periodogram, which follows a $\chi_2^2$ distribution. Then, similar to the above steps, through MCMC posterior predictive sampling, we can generate a large number of simulated periodograms. For each periodogram, we obtain its maximum outlier in residuals $T_R$. Comparing the observed values $T_R^{obs}$ with the simulated $T_R$ distribution, we can obtain the posterior predictive $p$-value.

Regarding broadband features, once again, employ the LRT for model selection. 
The null hypothesis corresponds to the noise model, while the alternative hypothesis corresponds to the noise model plus a QPO model. The QPO model is represented using a Lorentzian function
\begin{equation}
\label{eq3}
P(\nu) = \frac{A\gamma^2}{\gamma^2 + (\nu - \nu_0)^2},
\end{equation}
where $A$ is a normalization term, $\nu_0$ is the position of the peak, and $\gamma$ is half of the given full width at half maximum (FWHM). Similarly, we can generate a large number of simulated periodograms. For both the real periodogram and the simulated periodograms, we calculate the likelihood ratio using MAP estimation for both models. When comparing the real likelihood ratio with the distribution of the simulated likelihood ratios, we can obtain the posterior predictive $p$ values.

During the MCMC simulation, we employed flat priors for the model indexes, logarithmic priors for the model amplitude, and a normal prior with a mean of 2 and a width of 0.5 for white noise. The \texttt{emcee} and \texttt{stingray} packages were used for the primary computations described above~\cite{Foreman-Mackey:2012any,2019ApJ...881...39H}.

\section{Results}
For the periodogram of each GRB, we initially determine its noise model; we present the results in table~\ref{tab:1}. We choose the BPLC model when the $p$-value $<$ 0.05. Among the 525 GRBs, 146 are suitable for the BPLC model, while the remaining apply to the PLC model. The parameter values for each model corresponding to each GRB will serve as initial values for subsequent fitting. We show partial results of the QPO search in table~\ref{tab:2}. $f_0$ and $f_1$ represent the frequencies corresponding to the maximum residual value and the central frequency of the Lorentzian function fit, respectively. They represent the most likely frequencies for QPOs in narrowband and broadband features.

\begin{table}[tbp]
\centering
\begin{tabular}{lccr|ccc|ccccc}
\hline
\multirow{2}{*}{Pulse} & \multirow{2}{*}{LR} & \multirow{2}{*}{$p$-value} & \multirow{2}{*}{model} &
\multicolumn{3}{c}{PLC} &
\multicolumn{5}{|c}{BPLC} \\
\cline{5-7}
\cline{8-12}
\multicolumn{4}{c|}{} & $\beta$ & $\alpha$ & $\gamma$ & $\beta$ & $\delta$ & $\alpha_1$ & $\alpha_2$ & $\gamma$ \\

\hline
138 & 2.81 & 0.122 & PLC & 42.92 & 1.45 & 2.03 & 33.10 & 5.57 & 0.71 & 3.93 & 2.06 \\
185 & 15.11 & 0.020 & BPLC & 103.30 & 1.77 & 1.98 & 37.88 & 4.76 & 0.17 & 5.97 & 1.99 \\
206 & 5.31 & 0.047 & BPLC & 120.84 & 1.56 & 1.97 & 81.68 & 23.33 & 1.20 & 5.70 & 2.01 \\
207 & 38.48 & 0.000 & BPLC & 1287.93 & 1.48 & 1.94 & 73.33 & 30.72 & 0.27 & 3.34 & 2.11 \\
218 & 2.64 & 0.123 & PLC & 25.93 & 1.34 & 1.91 & 20.68 & 14.15 & 0.99 & 4.29 & 1.93 \\
289 & 7.60 & 0.010 & BPLC & 15.24 & 1.10 & 1.91 & 8.14 & 3.16 & -0.43 & 1.88 & 1.94 \\
373 & 0.00 & 0.709 & PLC & 44.11 & 1.74 & 1.96 & 84.64 & 33.14 & 1.72 & 1.74 & 1.96 \\
\hline
\end{tabular}
\caption{Results of the noise model fitting.\label{tab:1}}
\end{table}

\begin{table}[tbp]
\centering
\begin{tabular}{lccccc}
\hline
Pulse & $f_0$ & $T_R$ & $p$-value & $f_1$ & LR \\
\hline
138 & 466 & 15.1 & 0.800 & 712 & 12.0 \\
185 & 100 & 15.5 & 0.725 & 277 & 2.6 \\
206 & 814 & 18.9 & 0.224 & 313 & 5.1 \\
207 & 897 & 14.2 & 0.735 & 467 & 9.0 \\
218 & 130 & 18.6 & 0.221 & 550 & 9.3 \\
289 & 585 & 15.5 & 0.745 & 586 & 9.0 \\
373 & 208 & 19.1 & 0.147 & 261 & 17.6 \\
\hline
\end{tabular}
\caption{The results obtained through narrowband and broadband features.\label{tab:2}}
\end{table}

\begin{figure}[t]
\centering
\includegraphics[width=.45\textwidth]{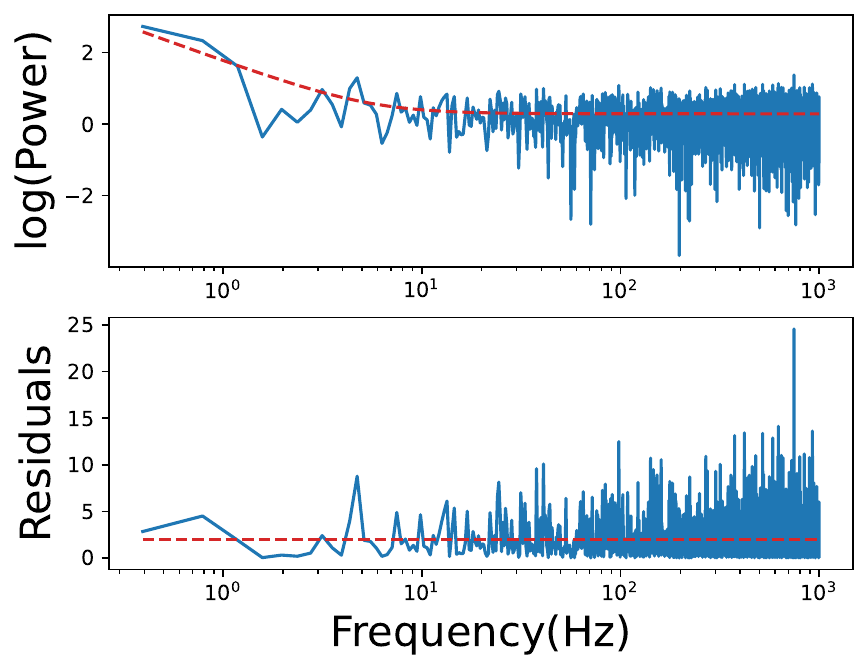}
\qquad
\includegraphics[width=.45\textwidth]{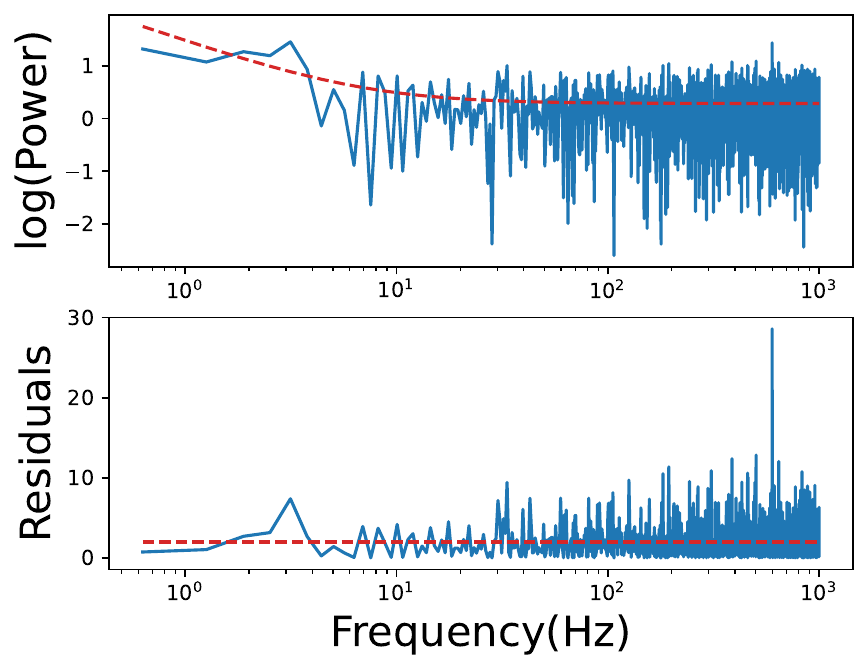}
\caption{Periodogram and residuals for BATSE pulse 2523 (left panel) and BATSE pulse 6439 (right panel). Pulse 2523 has the smallest $p$-value and pulse 6439 has the largest $T_R$. \label{fig:1}}
\end{figure}

We first examined the narrowband features. For each GRB, we recorded its maximum residual value $T_R$, corresponding frequency $\nu$, and the $p$-value obtained by comparing it with the simulated periodograms. No results exceeding 3$\sigma$ were found. 
The smallest $p$-value is 0.006 for pulse 2523, corresponding to a confidence level of 2.75$\sigma$. In figure~\ref{fig:1}, we provide results examples for pulse 2523 and pulse 6439, which have the smallest $p$-value and the largest $T_R$, respectively. The noise models for both GRBs are represented using the PLC model. The fitting results are illustrated by the red dashed lines in the periodograms. Although not very apparent in the periodograms, significant peaks can be observed in the residual plot. Pulse 2523 exhibits the maximum residual value $T_R$ at the frequency of 747 Hz, which is 24.6. Pulse 6439 has its $T_R$ at 599 Hz, which is 28.6. However, through Monte Carlo simulations of the periodograms for both bursts, their $p$-values are 0.006 and 0.016, respectively. Even the $p$-values are not trial-corrected, their significance does not exceed 3$\sigma$. 

\begin{figure}[t]
\centering
\includegraphics[width=.9\textwidth]{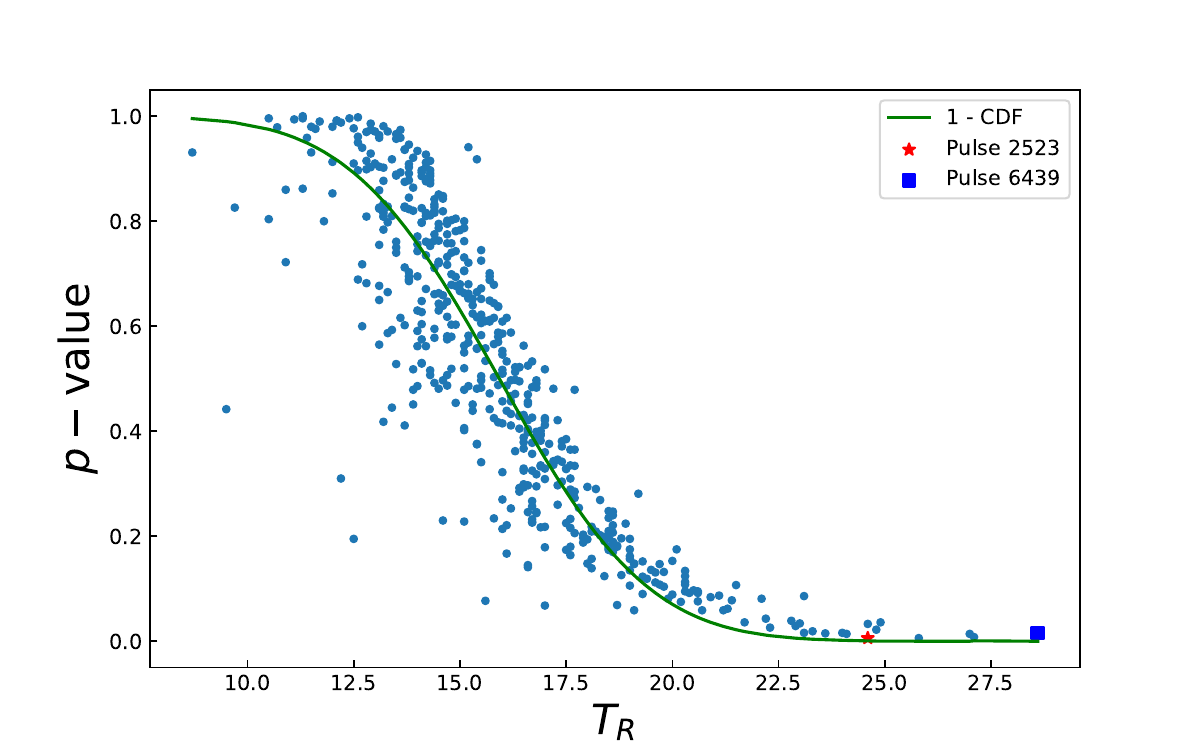}
\caption{Scatter plot of $p$-values with respect to $T_R$ for all GRBs. The red star represents BATSE pulse 2523 while the blue square represents pulse 6439. The green line represents 1 minus the theoretical cumulative distribution function of $T_R$.\label{fig:2}}
\end{figure}

To investigate broadband features, the primary challenge lies in the substantial computational load. In contrast to narrowband features, exploring broadband features introduces the Lorentzian function as the QPO model. This involves more than just adding a few parameters. To avoid the Lorentzian function from getting trapped in local optima as much as possible, at each frequency, we fix the centroid of the Lorentzian to that frequency and fit other parameters. This results in an increase in the computational load by a factor of N, where N is the number of data points in a periodogram. 

Simulating a large number of periodograms for model selection with each GRB data is quite challenging. Is it possible to selectively analyze only those GRBs with relatively large likelihood ratios in more detail? The answer is, to a certain extent, yes. In figure~\ref{fig:2}, we plot the scatter map of $p$-value with respect to $T_R$ for each GRB. It is easy to observe that the data can be fitted with a monotonically decreasing curve. Here, we fit a Gaussian function to the distribution of $T_R$ and calculate the theoretical cumulative distribution function (CDF) of the fitting result. The green line in the figure represents 1 minus this cumulative distribution function. It can be observed that the two match quite well. 

\begin{figure}[t]
\centering
\includegraphics[width=.9\textwidth]{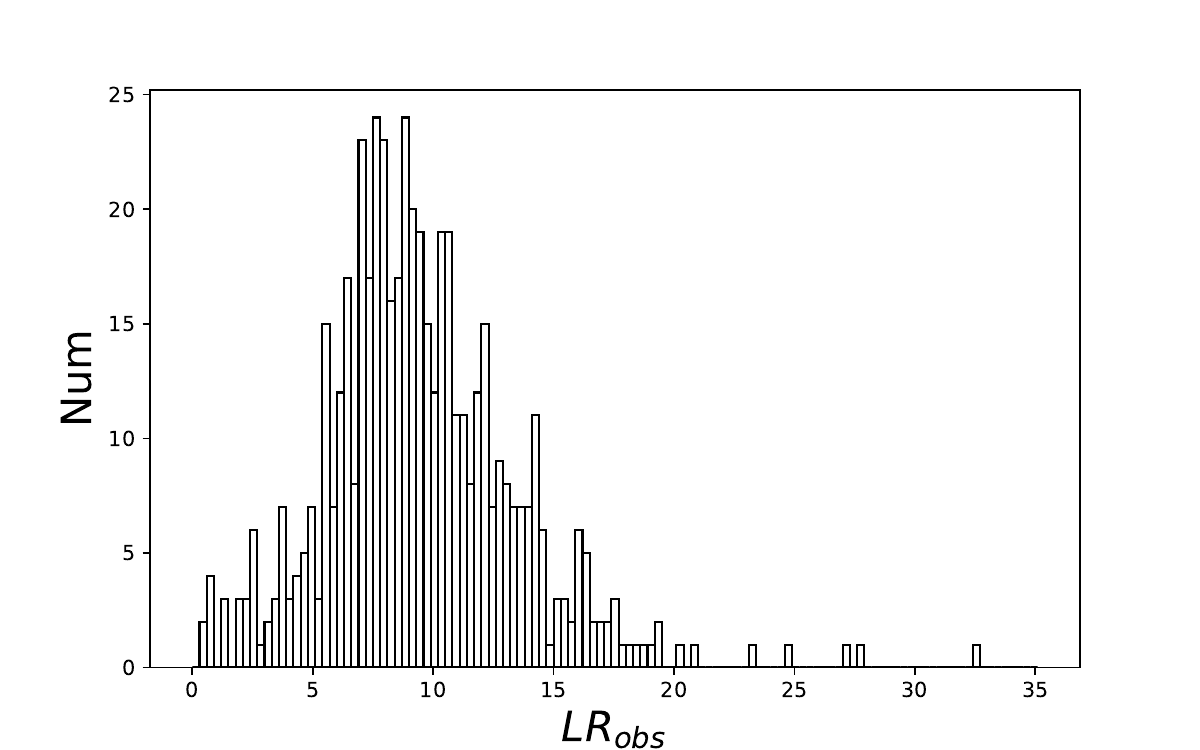}
\caption{Distribution of likelihood ratios for all GRBs. The null hypothesis is the noise model, and the alternative hypothesis is the noise model plus the QPO model.\label{fig:3}}
\end{figure}
That is to say, larger $T_R$ values often correspond to smaller $p$-values. 
We extend this conclusion to the exploration of broadband features because the two methods are almost the same, with the difference being the statistical measures they use as criteria. Therefore, we decide to calculate the likelihood ratio for each GRB and proceed with further analysis for those with larger values. The distribution of likelihood ratios for all GRBs is depicted in figure~\ref{fig:3}. We choose to conduct a simulation analysis for those with likelihood ratio values greater than 20, totaling 7 GRBs, with the following identifiers: 677, 2217, 2614, 3323, 6800, 7063, and 7427. Due to the complexity of the periodogram for 2614, we exclude it from further analysis. We perform 5000 Monte Carlo simulations for the periodogram of each remaining GRB, calculating the reference distribution for the likelihood ratio. The results are presented in table~\ref{tab:3}. The smallest $p$-value from a single trial is 0.0004 for pulse 7427. But a conservative estimate requires trial correction. Note that here we do not need to correct for the number of frequencies, as in both narrowband and broadband feature explorations, when calculating the reference distribution, we use the maximum residual value and the maximum likelihood ratio among all frequencies. We correct by the total number of GRBs in our sample, i.e., 525. Then, the smallest $p$-value increases significantly to 0.21, corresponding to a confidence level of 1.25$\sigma$, indicating that there is insufficient evidence to suggest the presence of QPO in these signals. 

\begin{table}[tbp]
\centering
\begin{tabular}{ccccc}
\hline
Pulse & $\nu_0$ & LR & $p$-value (single-trial)& $p$-value (trial-corrected)\\
\hline
677 & 59 & 23.3 & 0.0008 & 0.4400\\
2217 & 256 & 20.7 & 0.0035 & 1.8567\\
3323 & 271 & 20.1 & 0.0021 & 1.1242\\
6800 & 111 & 32.5 & 0.0070 & 3.6585\\
7063 & 143 & 24.6 & 0.0004 & 0.2106\\
7427 & 370 & 27.1 & 0.0004 & 0.2119\\
\hline
\end{tabular}
\caption{The results of the broadband search for six GRBs.\label{tab:3}}
\end{table}

\begin{figure}[t]
\centering
\includegraphics[width=.45\textwidth]{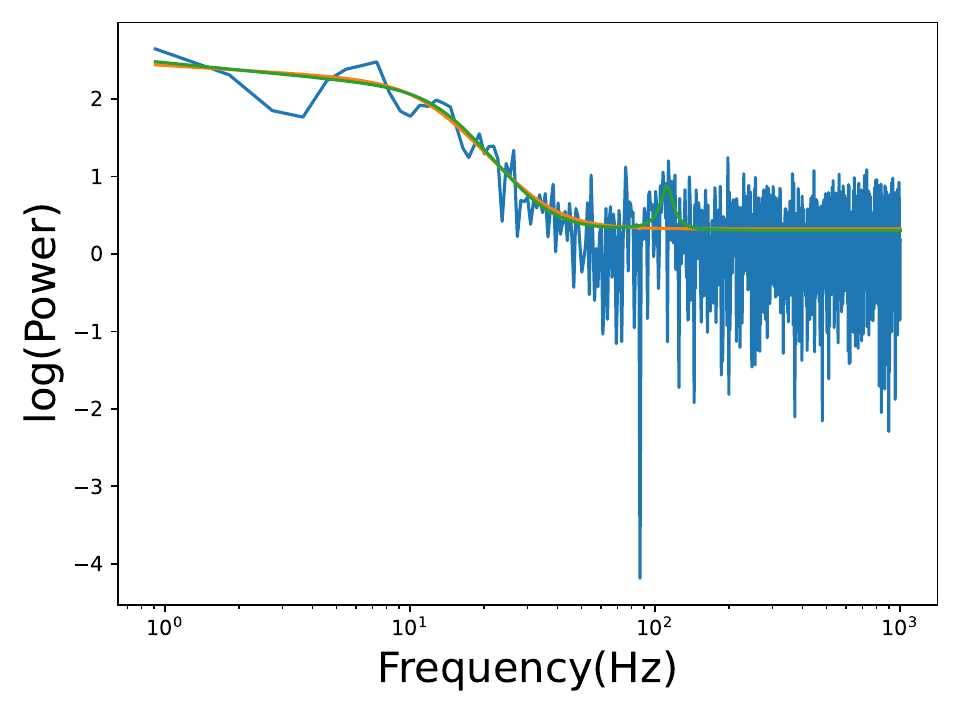}
\qquad
\includegraphics[width=.45\textwidth]{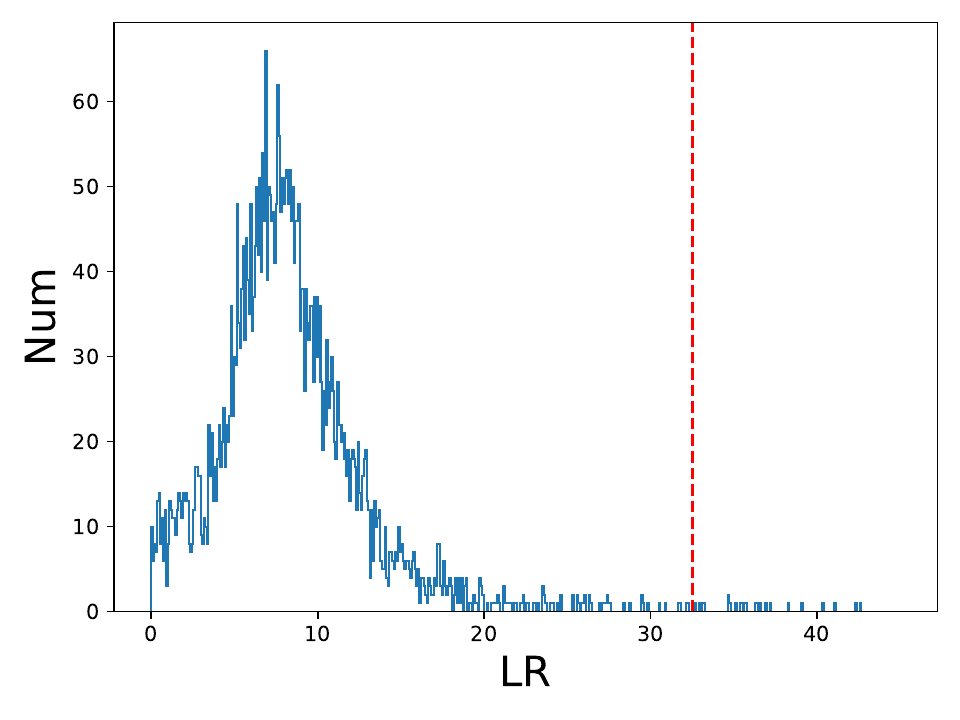}
\caption{Periodogram fitting and LRT for BATSE pulse 6800. In the left panel, the yellow line represents the fitting of the BPLC model, and the green line represents the fitting of the BPLC model with a Lorentzian function. In the right panel, the blue histogram represents the simulated reference distribution, and the red dashed line represents the observed likelihood ratio value.\label{fig:4}}
\end{figure}

In figure~\ref{fig:4}, we present the fitting results and LRT results for the periodogram of BATSE pulse 6800 as an example. For pulse 6800, the BPLC model is more suitable for describing its noise model than the PLC model. Among all GRBs, it has the largest likelihood ratio value. Simultaneously, in the figure, we can clearly see the distinct outline of a peak around 110 Hz. However, simulation results indicate that this is insufficient to be considered a QPO signal. Furthermore, although pulse 6800 has the largest likelihood ratio value among the 6 GRBs subjected to detailed analysis, its false positive rate is the highest. We speculate that this might be due to differences in the frequency resolution of the periodogram. Since the duration of each GRB data varies, with a shorter duration resulting in lower frequency resolution of the periodogram, pulse 6800, being the second shortest in the 6 GRBs, has fewer data points in its periodogram compared to the others. Fewer data points are more influenced by noise, and during Monte Carlo simulations to calculate the reference distribution, it is more likely to generate larger likelihood ratio values.

\section{Conclusion and Discussion}
In summary, our systematic QPO search on BATSE short burst data, considering both narrowband and broadband characteristics, did not yield compelling evidence. Despite some periodograms showing apparent signals, the likelihood of these being false positives is relatively high, especially after trial correction. No signals reached significance beyond the 3$\sigma$ threshold.

In astronomical problems analogous to target detection, a common framework is often employed: initially, a statistical measure is chosen as a criterion, followed by the calculation of the reference distribution for this statistic. Subsequently, the observed value is compared with the reference distribution to determine the false positive rate. This type of problem often be viewed as a nested significance test in the context of model selection. The challenge lies in determining the reference distribution, which may be known or unknown depending on the problem at hand. Traditionally, the $\chi^2$ distribution has been a frequent choice for the LRT, and the $F$-distribution has been utilized for the $F$-test in model selection scenarios~\cite{Band:1996re}. It was highlighted by~\cite{2002ApJ...571..545P} that for these tests and reference distributions to be applicable, the two models being compared should be nested, and the parameter values of the more complex model should not be on the boundary when it reduces to the simpler one. However, the QPO search precisely does not satisfy the second condition because both $T_R$, representing narrowband features, and the Lorentzian function, representing broadband features, regress to the simpler model when the amplitude is zero. And the amplitude being zero is exactly the lower limit of the parameters. In such a scenario, the reference distribution can only be determined through Monte Carlo simulations.

\begin{figure}[t]
\centering
\includegraphics[width=.9\textwidth]{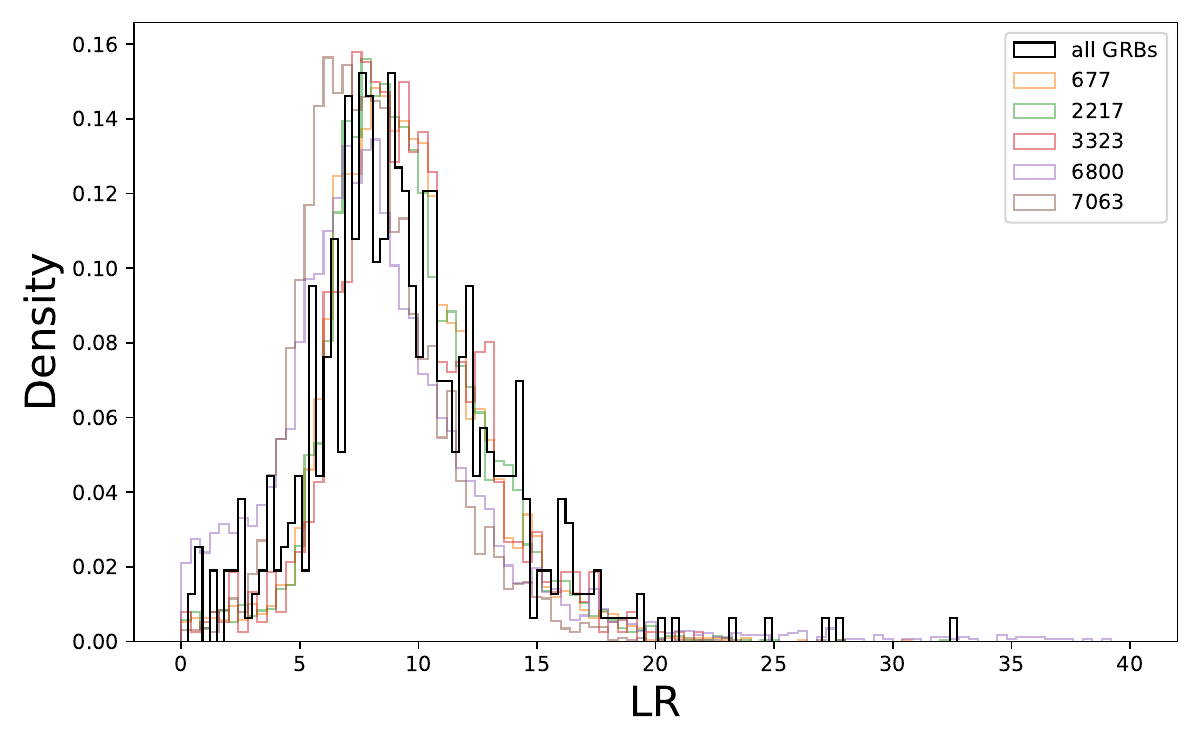}
\caption{Density plot of the likelihood ratio distribution. The black line, similar to figure~\ref{fig:3}, represents the likelihood ratio distribution for all GRBs, while the colored lines represent the reference distributions calculated through simulations for the five specific GRBs.\label{fig:5}}
\end{figure}

To overcome computational limitations, we computed the reference distribution for $T_R$ individually for each GRB in the case of narrowband features. For broadband features, we specifically selected those with higher likelihood values for constructing the reference distribution. This approach was motivated by the observed relationship between $T_R$ and $p$-value, where data statistics with greater extremes tended to yield more extreme $p$-values. From an alternate perspective, this implies that, for the statistic under examination, the reference distribution for each GRB is not significantly distinct from the overall distribution of all GRBs.
In figure~\ref{fig:5}, a comparison between the two distributions for LRT is presented. The black line illustrates the overall likelihood ratio distribution for all GRBs, while lines of different colors represent the reference distributions calculated for five specific GRBs: pulse 677, 2217, 3323, 6800, and 7063. Notably, these distributions exhibit a high degree of similarity, concentrated between 0 and 20, with a small probability of exceeding 20. The likelihood ratio reference distribution for pulse 6800 deviates the most from the overall distribution, displaying a relatively higher proportion exceeding 20. As mentioned earlier, this discrepancy could be influenced by various factors such as data length and resolution.

In the study of~\cite{Chirenti:2023dzl}, a significant contributing factor to the identification of two GRBs with kilohertz QPOs is the distinctive approach to $p$-value corrections. They link the reference distribution to the $T_{90}$ (the duration of bursts). Specifically, shorter $T_{90}$ durations are associated with a higher likelihood of having larger values in their reference distribution. Consequently, there is no necessity to apply a correction by multiplying it with the total number of samples in the dataset. This finding aligns with the outcomes observed in BATSE pulse 6800, where the reference distribution contains numerous larger values than others. This is attributed to the high-frequency resolution of its periodogram, resulting in the fewest data points within the 0-1000 Hz range. Unfortunately, even with a non-conservative estimate, the $p$-value for a single experiment in the simulations of pulse 6800 is 0.007, which still falls short of significance.

In future research endeavors, in addition to the pursuit of identifying QPOs, it may prove beneficial to thoroughly assess the significance of detected QPOs. Some recent studies have already delved into the impact of time series length and resolution on the accuracy of results. For instance, \cite{Huebner:2021aaa} demonstrated that in the case of non-stationary time series, the significance of QPOs might be overestimated. Another study by \cite{Li:2022jxj} explored the calculation of reference distributions using signals of varying lengths, uncovering diverse outcomes. As systematic searches often demand substantial computational resources, such insights can guide the selection of appropriate methods for period searches, potentially reducing the computational workload.

\acknowledgments

We thank Weihua Lei and Yu Liu for their helpful discussions.
This work is supported by the National Key R\&D Program of China (2022SKA0130100). The computation is completed in the HPC Platform of Huazhong University of Science and Technology.





\end{document}